\documentclass[journal=jctcce,manuscript=article,layout=traditional]{achemso}
\setkeys{acs}{doi = true, etalmode = firstonly, maxauthors = 15, abbreviations = true}
\usepackage{graphicx}
\usepackage{dcolumn}
\usepackage{bm}
\usepackage{verbatim}
\usepackage{soul}
\usepackage{float}
\usepackage{xcolor}
\usepackage{chemformula} 
\usepackage[T1]{fontenc} 
\usepackage{multirow}
\usepackage[fontsize=10pt]{fontsize}
\usepackage[colorlinks = true, allcolors = blue,]{hyperref}
\usepackage{orcidlink}

\title[\texttt{vdW-WanMBD}]{Dissecting van der Waals interactions with Density Functional Theory – Wannier-basis approach}

\author{Diem Thi-Xuan Dang\,\orcidlink{0000-0001-7136-4125}}
\email{dangt1@usf.edu}

\author{Dai-Nam Le\,\orcidlink{0000-0003-0756-8742}}
\email{dainamle@usf.edu}

\author{Lilia M. Woods\,\orcidlink{0000-0002-9872-1847}} 
\email{lmwoods@usf.edu (Corresponding author)}
\affiliation{Department of Physics, University of South Florida, Tampa, Florida 33620, USA}

\begin{document}

\begin{abstract}
A new scheme for the computation of dispersive interactions from first principles is presented. This cost-effective approach relies on a Wannier function representation compatible with density function theory descriptions. This is an electronic-based many-body method that captures the full electronic and optical response properties of the materials. It provides the foundation to discern van der Waals and induction energies as well as the role of anisotropy and different stacking patterns when computing dispersive interactions in systems. Calculated results for binding energies in layered materials, such as graphite, hBN, and \ch{MoS2} give encouraging comparisons with available experimental data. Strategies for broadened computational descriptions of dispersive interactions are also discussed. Our investigation aims at stimulating new experimental studies to measure van der Waals energies in a wider range of materials, especially in layered systems. 	

Published version is fount at Ref.\citenum{Dang2025}, Dang et al., \textit{Comp. Phys. Comm.} \textbf{310}, 109525 (2025).

\end{abstract}
\maketitle

\begin{figure}[H]
    \begin{center}
    \includegraphics[width = 0.9 \columnwidth]{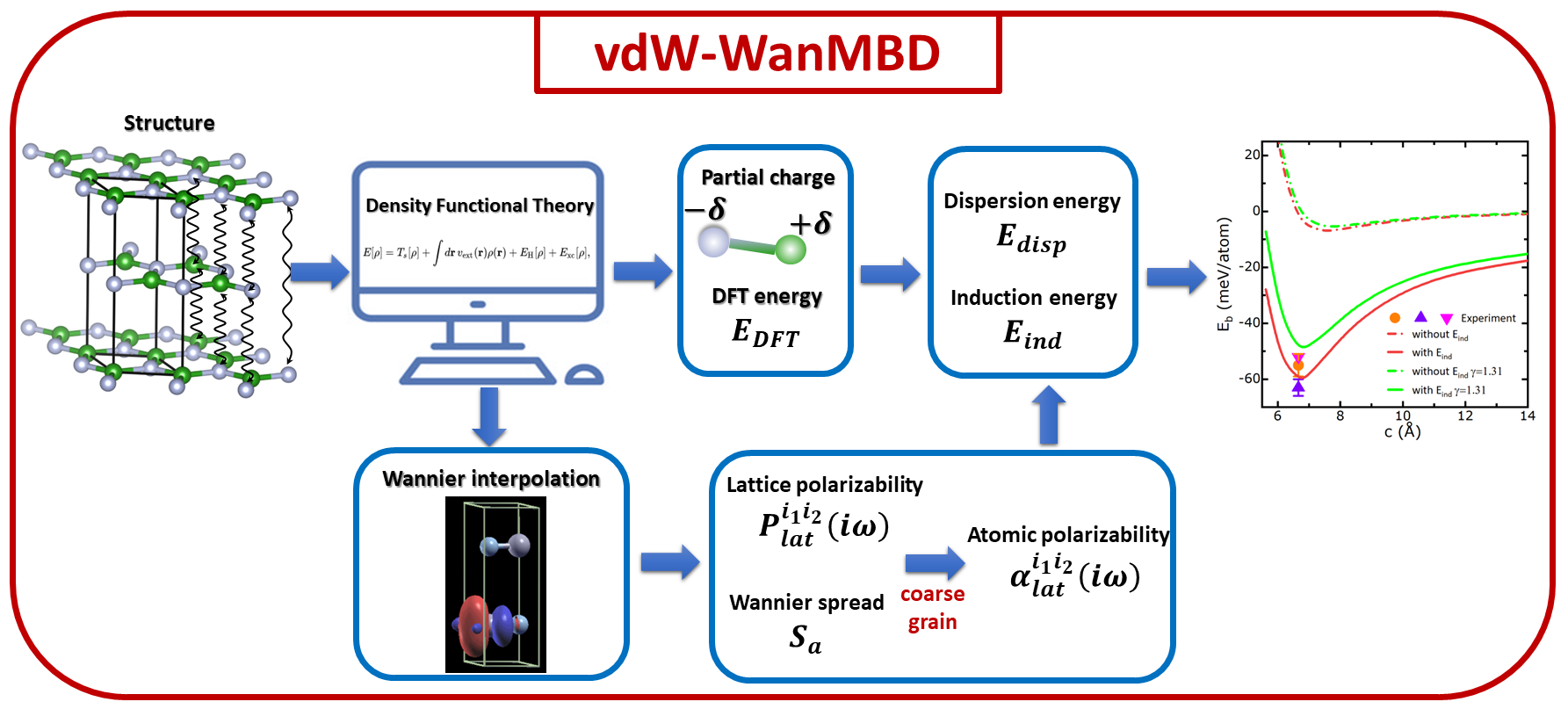}
    \end{center}
\end{figure}

\section{\label{sec:1}Introduction}

Layered materials are a driving factor for many discoveries as they can host limitless possibilities of new properties, tuning capabilities, and novel device applications. A unifying fundamental feature of such materials is the van der Waals (vdW) interaction responsible for keeping the individual chemically inert layers together in a stable system. Even though the vdW interaction is considered to be relatively weak due to its electromagnetic origin \cite{Ramalho2013, Hermann2017}, it plays a dominant role when stronger chemical forces are absent. The drive towards fundamental understanding of low dimensional and layered materials is closely related to being able to calculate vdW interactions in order to describe materials stability and stacking as well as a multitude of other properties.

Standard Density Functional Theory (DFT), a well-established computational tool, does not account for long-ranged charge correlations \cite{Ramalho2013, Grimme2016, Claudot2018}. The exact correlation energy for the vdW interaction obtained via the adiabatic-connection fluctuation-dissipation theorem (ACFD) \cite{Hermann2017, Stohr2019} cannot be computed exactly in systems beyond a few atoms, thus all currently available approaches rely on some approximations. Currently, quantum Monte Carlo (QMC) and Random Phase Approximation (RPA) that rely on DFT Kohn-Sham orbitals are perhaps the most accurate methods for vdW computations \cite{Foulkes2001,Nguyen2009}. However, the high computational cost and the inability to separate the long-ranged interaction from other short-ranged correlation terms make the application of such quantum chemical methods problematic. This has motivated the development of different approaches that are more computationally effective.

Atomistic methods that rely on the dipolar polarizabilities of the atoms enable calculations of vdW energies rather efficiently.  Many-body dispersion (MBD) computations utilize models by representing atoms in materials as harmonic oscillators coupled via dipolar interactions suggesting that charges are assumed to be rather confined to the atoms \cite{Dobson2023}. RPA implementations based on bare atomic polarizabilities and dipolar coupling are also available. Such methods are MBD-compatible and they compute the vdW energy based on density-density correlation functions \cite{Tkatchenko2013,Berland2015,Ambrosetti2016b,Ambrosetti2019b}. Non-local functionals for vdW energy computations have also been developed. They rely on different approximations related to the plasmon propagator properties and many of them also use external parameters \cite{Dion2004, Lee2010, Hamada2014}. 

The interplay of the atomic distribution in the materials, their polarization properties and underlying electronic structure makes the description of vdW interactions complex, which motivates new developments of first principles methods for more accurate and cost-effective calculations. Computing vdW energies based on correlation effects that capture the {\it full electronic structure} and {\it overall polarizability} of the system needs further efforts.  Other effects associated with materials anisotropy and their full optical response properties are also factors that need further evaluations in vdW computations from first principles. Less assumptions and removing external parameters in computational scheme are also needed for much improved reliability of the computations.

In this paper, we propose a computationally efficient method for dispersive interactions, that is based on maximally localized Wannier Function (WF) representation compatible with DFT simulations. The method enables distinction of various factors affecting the dispersive interactions, including optical anisotropy, polarity of atomic bonds, and stacking patterns in layered materials. Distinctions between the vdW and induction energies in the dispersive energies is enabled within a Kubo formalism that takes into account the electronic structure properties of the materials. Comparisons with experimental data are encouraging, although measurements with different techniques in a variety of materials are needed for further analysis of the complex nature of long ranged interactions and validation of computational methods and results.

\section{\label{sec:2}vdW method based on DFT-Wannier function interface: \texttt{vdW-WanMBD}}
\subsection{Perturbation Theory with Many-body Interactions}

The total energy of a given system can be expressed as 
\begin{eqnarray}
    \label{eqn:1}
    E_{total} = E_{DFT} + E_{disp}. 
\end{eqnarray}
where $E_{DFT}$ is the usual self-consistent DFT energy and $E_{disp}$ is the dispersion energy. The total DFT energy in the original Kohn-Sham theory is defined in terms of a functional of the electron density $\rho( \mathbf{r}) = \sum_{i=1}^N| \psi_i ( \mathbf{r})|^2$  for $N$ electrons occupying orbitals with corresponding wave functions $\psi_i (\mathbf{r})$ \cite{Hohenberg1964, Kohn1965}. It is given as 
\begin{eqnarray}
    \label{eqn:2}
    E_{DFT} [\rho] = T[\rho] + E_{ext} [\rho] + E_H [\rho] + E_{xc} [\rho]       
\end{eqnarray}
where $T[\rho]$ is the kinetic energy, $E_{ext} [\rho]$ is the external energy, $E_H [\rho]$ is the Hartree energy, and $E_{xc} [\rho]$ is the exchange-correlation energy. Standard DFT uses different approximations to compute the unknown $E_{xc} [\rho]$. Most are based on local density approximation (LDA), generalized gradient approximation (GGA), and hybrid exchange-correlation functionals for more accurate calculations. These approximations, however, consider only local properties in the materials and do not include density fluctuations associated with electron-electron correlations \cite{Klimevs2012, Berland2015, Woods2016}. These missing long-ranged nonlocal effects give rise to the vdW dispersion interaction, which is quite important when considering systems with chemically inert components. The vdW dispersion energy $E_{disp}^{vdW}$ between two atomistic systems A and B can be obtained by treating their electron-electron Coulomb interaction within perturbation theory captured by the nonlocal Hamiltonian \cite{Stone2013}
\begin{eqnarray}
    \label{eqn:3}
     && \hat{H}_{nl} = \sum_{u \in A, v \in B} \hat{H}_{nl}^{uv} = \sum_{u \in A, v \in B} \left[- \sum_{i_1, i_2}^{x, y, z} T_{uv}^{i_1 i_2 } \hat{\mu}_u^{i_1} \hat{\mu}_v^{i_2} - \frac{1}{3} \sum_{i_1,i_2, i_3}^{x, y, z} T_{uv}^{i_1 i_2 i_3} \left( \hat{\mu}_u^{i_1} \hat{\Theta}_v^{i_2 i_3} - \hat{\Theta}_u^{i_1 i_2} \hat{\mu}_v^{i_3} \right) \right. \nonumber \\
    && - \frac{1}{45} \sum_{i_1,i_2,i_3,i_4}^{x,y,z} T_{uv}^{i_1 i_2 i_3 i_4} \left( 3 \hat{\mu}_u^{i_1} \hat{\Omega}_v^{i_2 i_3 i_4} - 5 \hat{\Theta}_u^{i_1 i_2} \hat{\Theta}_v^{i_3 i_4 } + 3 \hat{\Omega}_u^{i_1 i_2 i_3} \hat{\mu}_v^{i_4} \right) \nonumber \\
    && + \frac{1}{45} \sum_{\substack{i_1, i_2,i_3\\i_4,i_5}}^{x,y,z} T_{uv}^{ i_1 i_2 i_3 i_4 i_5} \left( \hat{\theta}_u^{i_1 i_2} \hat{\Omega}_v^{i_3 i_4 i_5} - \hat{\Omega}_u^{i_1 i_2 i_3 } \hat{\Theta}_v^{i_4 i_5} \right)  \left. - \frac{1}{225} \sum_{ \substack{i_1, i_2, i_3\\ i_4, i_5, i_6}}^{x, y, z} T_{uv}^{i_1 i_2 i_3 i_4 i_5 i_6} \hat{\Omega}_u^{i_1 i_2 i_3} \hat{\Omega}_v^{i_4 i_5 i_6 } + \ldots \right] ,   
\end{eqnarray}
where $\hat{\mu}_u^{i_1}$, $\hat{\theta}_u^{i_1 i_2}$, $\hat{\Omega}_u^{i_1 i_2 i_3}, \ldots$ are the components of the dipole, quadruple, octupole, ... operator moments of atoms $u$ with $i_1,i_2,i_3, \ldots = x, y, z$ components. The above Hamiltonian is for materials with charge neutrality, and it takes into account long-ranged interactions between electrons residing in atom $u$ located at $\mathbf{R}_u$ atomic site of material A and atom $v$ located at $\mathbf{R}_v$ of material B captured by the interaction tensor 
\begin{eqnarray}
    \label{eqn:4}
    T_{uv}^{i_1 i_2 i_3 \ldots} = \dfrac{1}{4 \pi \varepsilon_0} \nabla_{i_1} \nabla_{i_2} \nabla_{i_3} \ldots \left( \dfrac{1}{R_{uv}} \right)  ,
\end{eqnarray}      
where $R_{uv} = |\mathbf{R}_u - \mathbf{R}_v |$ is the separation between two atomic sites. 

By retaining only the first dipole-dipole term in $H_{nl}$, the vdW interaction energy can be written as \cite{Ambrosetti2014c,Stohr2019}
\begin{eqnarray}
    \label{eqn:5}
    E_{disp}^{vdW} = - \sum_{n = 2}^\infty \dfrac{(-1)^n}{n}\int_0^\infty \dfrac{d\omega}{2 \pi}  \mathrm{ Tr } \left\{ \left[ \sum_{\substack {u \in A \\ v \in B}} \boldsymbol{\alpha}_u ( i \omega) \mathbf{T}_{uv} (\mathbf{R}_{uv} ) \right]^n \right\} .  
\end{eqnarray}
The above expression is consistent with RPA and it is obtained by employing a Dyson equation for point-like polarizabilities $\boldsymbol{\alpha}_u (i \omega)$ at the atomic sites $\mathbf{R}_u$ \cite{Ambrosetti2014, Stohr2019}. The polarizabilities are taken in the imaginary frequency domain as implied by the Casimir-Polder identities, which are also used in the derivation of Eq. \eqref{eqn:5} \cite{Casimir1948, Stone2013}. The dispersion energy $E_{disp}^{vdW}$ includes many-body interactions arising from dipolar correlations. We note that Eq. \eqref{eqn:5} can also be obtained within the coupled quantum harmonic oscillators method where each atom is treated as a single fluctuated dipole described by its dipolar polarizability $\alpha_u^{i_1 i_2} (i \omega)$ \cite{Kim2007, Shtogun2010, Le2024}. 
The dispersion energy is now recast in reciprocal space using periodic boundary conditions given that the atomic sites can be written as $\mathbf{R}_{u,v} = \mathbf{r}_{a,b} + \mathbf{L}_{u,v}$, where $\mathbf{r}_{a,b}$ are the atomic positions in the unit cell of the materials, while $\mathbf{L}_{u,v}$ are their lattice translation vectors. It can then be written as
\begin{eqnarray}
    \label{eqn:6}
    E_{disp}^{vdW} = \sum_{\mathbf{k} \in IBZ} w_{\mathbf{k}} \int_0^{\infty} \dfrac{ \hbar d \omega}{2\pi} \mathrm{ Tr } \left\{ \ln \left[ 1 - \mathbf{A} (i \omega) \mathbf{T} (\mathbf{k}) \right] \right\},
\end{eqnarray}
where the summation is over the wave vector $\mathbf{k}$ in the first Brillouin zone (IBZ) with weight factors $w_\mathbf{k}$ corresponding to the symmetry degeneracy of each k-point \cite{Ambrosetti2021c}. The overall optical response of the interacting materials captured in $ \mathbf{A} (i \omega)$ (discussed further below) is expressed in imaginary frequency as obtained in the Adler-Wiser expression \cite{Adler1962, Wiser1963}.

Eq. \eqref{eqn:6} takes advantage of relating the dipole operator components in the first Brillouin zone $T_{ab, \mathbf{L}}^{i_1 i_2} = \int\limits_{IBZ} T_{ab}^{i_1 i_2} ( \mathbf{k}) \exp{\left(i \mathbf{k} \cdot \mathbf{L}\right)} d^3 \mathbf{k}$ to its Fourier components $T_{ab}^{i_1 i_2} (\mathbf{k}) = \sum\limits_{\mathbf{L} \neq \pmb{0}} T_{ab, \mathbf{L}}^{i_1 i_2 } \exp{\left(-i \mathbf{k} \cdot \mathbf{L}\right)}$  by exploiting the periodicity of the system through the translation vectors $ \mathbf{L} = \mathbf{L}_v - \mathbf{L}_u$. The interaction tensor $\mathbf{T}( \mathbf{k})$ is a $3N_{at} \times 3N_{at}$ ( $N_{at}$ – number of atoms in the cell) matrix whose elements are nonzero when $a$, $b$ indexes belong to materials A, B respectively, and they are zero otherwise. The elements of the interaction tensor also satisfy the relations $[ \mathbf{T}(- \mathbf{k})]_{ab}^{i_1 i_2} = {[ \mathbf{T}( \mathbf{k})]_{ab}^{i_1 i_2}}^* = [ \mathbf{T}( \mathbf{k})]_{ba}^{i_2 i_1}$. Eq. \eqref{eqn:6} is a consequence of the assumption that the polarizability of the interacting materials is only dependent on the frequency \cite{Ambrosetti2014, Buvcko2016, Poier2023}. One further takes that $A_{ab}^{i_1 i_2} (i \omega) = \alpha_a^{i_1 i_2} (i \omega) \delta_{ab}$, meaning that the components of the polarizability tensor are determined by the polarizabilities of the atomic sites in the unit cell. 

Eq. \eqref{eqn:6} is consistent with the ACFD theorem in which the unknown exchange-correlation kernel of the coupled system is neglected. It is also consistent with the RPA vdW interaction within the effective medium description, in which its kernel $\chi_0 ( \mathbf{R}_u, \mathbf{R}_v, i\omega) \mathcal{V} ( \mathbf{R}_u, \mathbf{R}_v )$ (here $\chi_0 ( \mathbf{R}_u, \mathbf{R}_v, i \omega)$ is the polarization function and $ \mathcal{V} ( \mathbf{R}_u, \mathbf{R}_v)$ is the Coulomb potential) is replaced with the equivalent from the MBD description term $ \mathbf{A}(i \omega) \mathbf{T} (\mathbf{k})$ \cite{DiStasio2014}. We further note that Eqs. \eqref{eqn:5} and \eqref{eqn:6} are equivalent to dispersion energy obtained by the coupled discrete dipole and the RPA models, as also shown in Ref. \citenum{Tkatchenko2013}. Let us also point out that the fluctuation-induced interactions in the microscopic Eq. \eqref{eqn:6} happen instantaneously, which is appropriate for separations at the nanoscale. At larger separations, retardation effects need to be considered, which is usually considered within the macroscopic (effective medium) Dzyaloshinskii-Lifshitz-Pitaevskii method \cite{Dzyaloshinskii1961}. 

\subsection{van der Waals Dispersion Energy Calculations with a Wannier Function Interpolation Scheme}

As discussed previously, the dispersion energy in all available computational methods essentially relies on some form of Eq. \eqref{eqn:6}. The differing features are related to how $E_{disp}^{vdW}$ is implemented in the ab initio simulations. The approach we present here relies on post-processing calculations of the polarizabilities and dispersion energy within a WF interpolation scheme, presented in detail in what follows.

We first begin by computing the electronic structure of the materials with standard DFT by using the Perdew-Burke-Ernzehof (PBE) exchange-correlation functional as implemented in the \textsf{Quantum ESPRESSO} package \cite{Giannozzi2009}. The electronic structure is then recast within the WF interpolation technique \cite{Wang2006}, in which the DFT wave functions are projected on conveniently chosen maximally localized WFs per atom \cite{Mostofi2008, Pizzi2020}. In this interpolation process Bloch-like states are defined, such that $\left| u_{n\textbf{k}}^{(W)} \right > = \sum_{\mathbf{R}} \exp{ \left[-i \mathbf{k} \cdot ( \mathbf{r} - \mathbf{R} - \boldsymbol{\tau}_n ) \right] } \left| \mathbf{R}n \right>$ where $n = 1, \ldots, M$ labels the band index for $M$ WFs per unit cell and $\mathbf{R}$ labels the unit cell. The characteristic features are the WF centers, given by $\boldsymbol{\tau}_n = \left< \mathbf{R}n \right| \mathbf{\hat{r}} \left| \mathbf{R}n \right>$ with $\hat{\mathbf{r}}$ being the position operator, and WF spreads $S = \sqrt{ \left< \mathbf{R}n \right| \hat{r}^2 \left| \mathbf{R}n \right> - \left< \mathbf{R}n \right| \mathbf{\hat{r}} \left| \mathbf{R}n \right>^2 }$. The DFT Hamiltonian, and its eigenstates and eigenenergies are then mapped into their equivalents in the WF representation \cite{Marzari1997, Azpiroz2018}. This alternative localized basis with its eigenstates and eigenenergies constitutes the basic tool of our subsequent computations.

The response properties of the atoms here are parameterized by coarse-graining the overall polarizability of the lattice $P_{lat}^{ i_1 i_2}$ into atomic fragments, such that
\begin{eqnarray}
    \label{eqn:8}
    \alpha_a^{i_1 i_2} (i \omega) = \dfrac{S_a^3 N_a}{ \sum\limits_{b \in UC} S_b^3 N_b } P_{lat}^{i_1 i_2} (i \omega) \Omega,  
\end{eqnarray}
where $S_a^3$ is the cubed largest WF spread of atom $a$ and $N_a$ is the number of WFs of atom $a$. This is a sensible representation of the atomic volume since the spread of each WF is associated with the spatial extension of the electrons associated with the atom. The weight factor $\sum\limits_{b \in UC} S_b^3 N_b$  sums over the product of the cubed WF spreads $S_b^3$ and the number of WFs $N_b$ associated with all atoms in the unit cell. One notes that the polarizability of the lattice $P_{lat}^{i_1 i_2} (i \omega)$  is obtained per volume, thus the total polarizability of all atoms in the unit cell is found as  $P_{lat}^{i_1 i_2} (i \omega) \Omega$, where $\Omega$ is the unit cell volume.

Note that Eq. \eqref{eqn:8} is different from previous studies utilizing the WFs approach for calculating vdW interactions. Specifically, in Refs. \citenum{Ambrosetti2012, Silvestrelli2013, Silvestrelli2019} the electronic charge partitioning are also represented by the WF spreads, however, the atomic polarizability is assumed to be proportional to the cube of WF spreads $\alpha_a (i \omega) \propto \gamma S_a^3$, where the constant $\gamma$ is set up by imposing the exact value for the H atom polarizability $\alpha_H = 4.5$ a.u for which $S_H = \sqrt{3}$ a.u. Such a construction is quite convenient, but it still relies on atomic dipoles modeled as harmonic oscillators. Ultimately, the vdW energy takes $C_6 R^{-6}$, $C_8 R^{-8}$ ,… expansion and the WF functions are conveniently used to compute the Hamaker constants $C_6$, $C_8$, ….

Eq. \eqref{eqn:8} is also different from other MBD methods, where the atomic polarizabilities are obtained within DFT. For example, the range-separated self-consistently screened (rsSCS) variant of the  MBD approach \cite{Tkatchenko2013, Ambrosetti2014, Buvcko2016} assumes that $\alpha_a^{i_1 i_2} (i \omega)$ is described by a single isotropic dipolar model  satisfying a self-consistent equation determined by the short-range interaction tensor $T_{SR}$ and the polarizability of Hirshfeld’s atom-in-molecule model $\alpha_a^{TS} (i \omega)$. This polarizability $\alpha_a^{TS}$  is evaluated using free-atom reference data with the self-consistent procedure based on the DFT electron density to account for short-range exchange-correlation and hybridization effects in the individual atomic polarizabilities. The fractionally ionic (FI) approach in MBD@rsSCS/FI improves the description of the atomic polarizabilities by linking the ionic charge and the frequency-dependent atomic polarizability \cite{Gould2016}. In the MBD-NL model \cite{Hermann2020}, the atomic dynamic polarizabilities are obtained by partitioning the Vydrov-Van Voorhis (VV) polarizability density with Hirshfeld weights where the VV polarizability is a semilocal functional of the electron density that models the local dynamic polarizability density. One notes that a common feature to all of these methods is that all are based on the harmonic oscillator representation of isotropic atomic dipoles.   

The model presented here, however, relies on the overall lattice polarizability tensor, which is  computed using WFs representation. While the WF charge partitioning is  applicable to metals, the localization of bare polarizabilities in MBD methods may not be the best choice in the presence of delocalized electrons (see, for example, Ref. \citenum{Dobson2023}). The polarizability tensor is found from the linear response Kubo formula involving optical transitions allowed by the selection rules for the dipole moment operator $\hat{\boldsymbol{\mu}} = e \hat{\mathbf{r}}$ \cite{Zhu2021, Popescu2020}. The components are expressed in terms of the WFs as
\begin{eqnarray}
    \label{eqn:9}
    P_{lat}^{i_1 i_2} (i \omega) = \dfrac{2 e^2}{\Omega} \sum_{\mathbf{k}\in IBZ} w_{\mathbf{k}} \sum_{m, n, m \neq n} \dfrac{(f_{n\mathbf{k}} - f_{m \mathbf{k}}) \epsilon_{n \mathbf{k},m \mathbf{k}}}{\epsilon_{n \mathbf{k}, m \mathbf{k}}^2 + (\hbar \omega)^2 } \mathcal{A}_{mn}^{i_1} ( \mathbf{k}) \mathcal{A}_{nm}^{i_2} (\mathbf{k}),  
\end{eqnarray}
where $\epsilon_{n \mathbf{k}, m \mathbf{k}} = \epsilon_{n \mathbf{k}} - \epsilon_{m \mathbf{k}}$ with $\epsilon_{n \mathbf{k}}$, $\epsilon_{m \mathbf{k}}$ being the eigenergies and $w_\mathbf{k}$ - the k-point weight capturing the symmetry degeneracy of each k-point in IBZ. Also, $\mathcal{A}_{mn}^{i_1} (\mathbf{k}) = i \left< u_{m \mathbf{k}} \right| \dfrac{\partial}{\partial k_{i_1}} \left| u_{n \mathbf{k}} \right>$ are the components of the Berry connection arising from the dipole moment operator components given as $ \left< m \mathbf{k} \right| \hat{\mu}_{i_1} \left |n \mathbf{k}_1 \right> =$ $e \left\{(1 - \delta_{mn} ) \delta ( \mathbf{k} - \mathbf{k}_1 ) \mathcal{A}_{mn}^{i_1} (\mathbf{k}) \right.$ $\left. + \delta_{mn} \left[ \delta( \mathbf{k} - \mathbf{k}_1 ) \mathcal{A}_{mn}^{i_1} (\mathbf{k}) + i \dfrac{\partial}{\partial \mathbf{k}} \delta( \mathbf{k} - \mathbf{k}_1 ) \right] \right\}$ in the Bloch picture with $\left| m \mathbf{k} \right>$, $\left| n \mathbf{k}_1 \right>$ denoting the Bloch states and $f_{n \mathbf{k}}$, $f_{m \mathbf{k}}$ are the Fermi distribution functions taken at 0 K temperature in the calculations \cite{Sipe2000}. The Berry connection components are computed with a WFs basis procedure in Ref. \citenum{Azpiroz2018}, as implemented in \textsf{WANNIER90} code.

The dipole tensor $\mathbf{T}( \mathbf{k})$ captures the atomistic arrangement of the interacting materials, which becomes divergent at very small separations. To remedy such non-physical behavior, the dipole tensor is modified by a short-range damping function $f(r_{ab, \mathbf{L}})$. This is a standard practice in all (semi)empirical and MBD computational schemes \cite{Buvcko2016, Grimme2016}. Here, we use the Fermi damping function $\bar{f}$ \cite{Wu2002, Grimme2004}, which modifies the interaction tensor,
\begin{eqnarray}
    \label{eqn:10}
    [T ( \mathbf{k})]_{ab}^{i_1 i_2} && = \sum_{ \mathbf{L} \neq 0 \text{ when } a = b} \bar{f}( r_{ab, \mathbf{L}} ) T_{ab, \mathbf{L}}^{i_1 i_2} \exp{\left(-i \mathbf{k} \cdot \mathbf{L} \right)}, \\
    \label{eqn:11}
    \bar{f} (r_{ab, \mathbf{L} }) && = \dfrac{1}{ 1 + \exp \left\{ - \left[ \frac{r_{ab, \mathbf{L}}}{ \gamma \left( S_a + S_b \right)} - 1 \right] \right\}  } , 
\end{eqnarray}
where $S_a + S_b$ is the sum of the \textcolor{blue}{largest} WF spreads for atoms $a$, $b$ and $\mathbf{r}_{ab, \mathbf{L}} = \mathbf{r}_a - \mathbf{r}_b + \mathbf{L}$. The coefficient $\gamma$ is an input parameter. Here we take $\gamma = 1.31$ as suggested in previous vdW calculations utilizing WFs \cite{Ambrosetti2012, Silvestrelli2013, Silvestrelli2016, Silvestrelli2019, Silvestrelli2020}. Such a value corresponds the ratio of the vdW radius and Wannier spread of isolated hydrogen atom at its ground state. Also, in previous studies the scaling denominator of $\mathbf{r}_{ab, \mathbf{L}}$ was assumed as the sum of the vdW radii of the atoms taken as input parameters. In the DFT-D and DFT-D2 methods, for example, the vdW radii are taken as the radius of the $0.01$ a.u. electron density contour of the atoms in the ground state within a Hartree-Fock type approximation scaled by numerical factors \cite{Grimme2004, Grimme2006}. In the TS method, the vdW radii are computed by rescaling the free-atom vdW radii for rare-gas atoms taken as one-half of the separation between the atoms in a relaxed dimer \cite{Tkatchenko2009}. Taking $\gamma=1$, on the other hand, corresponds to no external parameters in the damping function. In this case, Eq. \eqref{eqn:11} relies exclusively on the properties from the underlying lattice and electronic structures captured through the WFs basis interpolation.

\begin{figure}[H]
    \begin{center}
    \includegraphics[width = 0.75 \columnwidth]{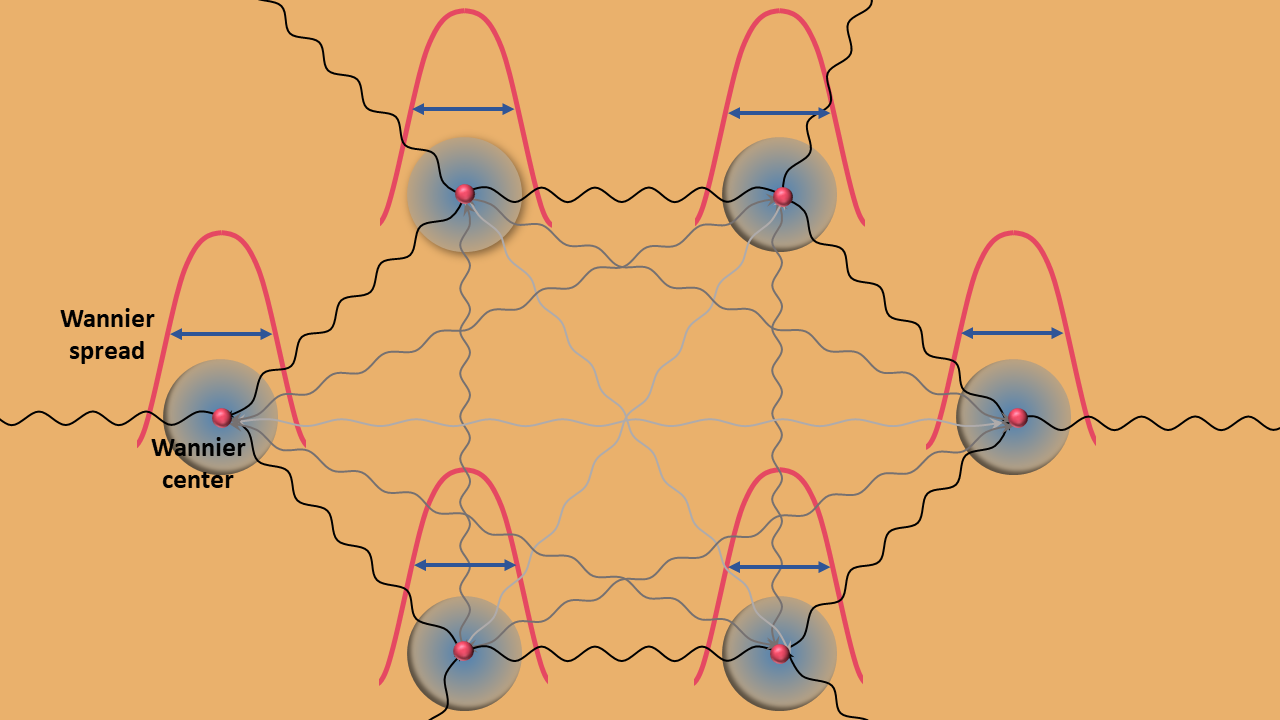}
    \caption{\label{fig:1} Schematic illustration of the \texttt{vdW-WanMBD} model, where atoms are represented by WF spreads and centers. }
    \end{center}
\end{figure}

In a nutshell, the entire computational scheme of the dispersion energy is now based on a WF  representation capturing the atomistic nature, electronic structure, and optical response of the materials (schematics in Fig. \ref{fig:1}). From the interpolation of the DFT electronic structure with the \textsf{WANNIER90} code, one obtains the maximally localized WFs whose centers and spreads can also be computed. Following the linear response Kubo formula, the polarizability tensor components of the entire system are found as a function of frequency taking into account the underlying electronic structure and atomistic nature of the interacting materials. The interaction tensor is calculated as a simple numerical integration, where the damping function determined by the WF spreads eliminates the divergences at close separations. 

We note that the dispersion energy and optical response discussed above are consistent with MBD and semiempirical pairwise models. However, there are several distinct features between the computational method described here and other available schemes. In most MBD approaches, the polarizabilities are usually approximated by isotropic dipolar oscillators with their screened interaction estimated through a self-consistent equation, while the dispersion energy is calculated separately and added to the DFT total energy \cite{Buvcko2016, Gould2016}. In the TS method, the polarizabilities are assumed to be proportional to free-atom values, where the proportionality constants are estimated using a Hirshfield partition of the charge density \cite{Tkatchenko2009}. The TS$+$SCF scheme assumes Drude-like polarizability with parameters obtained from a self-consistent equation \cite{Tkatchenko2012}. The density-dependent self-consistent (dDsC) procedure is similar to the TS and TS$+$SCF methods but with a different partition scheme for the charge density \cite{Steinmann2011}. In summary, the effect from the charge density of the materials is taken self-consistently assuming isotropic dipolar oscillators for the atoms, while the damping function is controlled by external parameters. The nonlocal density functionals, on the other hand, are the RPA family of approaches, where the density response in the ACFD formula is expressed through a nonlocal functional dependent on the density and its gradients. The earliest version assumes a plasmon-pole model for the density-density correlation function \cite{Dion2004}. The strong repulsion at smaller separations in the earlier nonlocal vdW models was improved by introducing various exchange-correlation factors and switching functions that are controlled with one or more input parameters fits based on available databases \cite{Berland2015, Chakraborty2020}. 

It appears that while various levels of self-consistency relying on the DFT density correlations are considered, the MBD  variants rely on different assumptions accompanied by adjustable parameters, such as scaling factors in the damping functions, and external parameters, such as atomic polarizabilities, vdW radii, and $C_6$ Hamaker constants of free atoms \cite{Buvcko2016, Hermann2020}. The approach presented here is a post processing computation of both the optical response of the system and its dispersion energy. This is an entirely electron-based approach employing WFs basis interpolation. It naturally considers the underlying electronic structure of the interacting materials while calculating all components of the polarizability tensor (Eq. \eqref{eqn:9} without assuming isotropic dipolar oscillators (as done in the MBD methods) or restricting the response due to isotropic density-density correlation functions (as done in the nonlocal density methods). Taking the scaling factor $\gamma=1$ in Eq. \eqref{eqn:10} essentially corresponds to no adjustable parameter in the model. Our approach is also much less computationally demanding and it builds on earlier WF schemes, such as vdW-WF2, vdW-WF2s, and vdW-WF2-x, which rely on static isotropic atomic polarizabilities assumed to be proportional to atomic volumes expressed in terms of WF spreads \cite{Ambrosetti2012, Silvestrelli2013, Silvestrelli2019}. These vdW-WF methods, however, only compute Hamaker constants associated with the $R^{-6}, R^{-8}$ scaling law from the change in zero-point energy of coupled dipole-dipole oscillations while higher many-body contributions to the vdW energy are not considered.

\subsection{Induction Energy Calculations}
Dispersive interactions for polar materials contain not only vdW contributions originating from dipoles due to the electromagnetic fluctuations, but also from the long-ranged coupling of their permanent dipoles due to uneven charge distributions. The induction energy can be obtained from the Hamiltonian $ \hat{H}_{ind} = \sum\limits_{u,v} \sum\limits_i^{x,y,z} T_{uv}^i  \left( q_u \hat{\mu}_v^i - q_v \hat\mu_u^i \right)$, where $q_{u,v}$ are the partial atomic charges \cite{Smith2012, Stone2013}. Within second order perturbation theory,  we find  $E_{disp}^{ind} = -\frac{1}{2} \sum\limits_{u,v} \sum\limits_{i_1, i_2}^{x,y,z} T_{uv}^{i_1} T_{uv}^{i_2} \left[ q_u^2 \alpha_v^{i_1 i_2} (0) + q_v^2 \alpha_u^{i_1 i_2} (0) \right]$, where $\alpha_{a,b}^{i_1 i_2} (0)$ are the polarizabilities at zero frequency. Utilizing periodic boundary conditions, the induction energy per unit cell is then found as
\begin{eqnarray}
    \label{eqn:12}
    E_{disp}^{ind} = - \frac{1}{4} \frac{1}{(4 \pi \varepsilon_0)^2}  \sum_{a \in A, b \in B}^{UC} \sum_{i_1, i_2}^{x,y,z} \left[ q_a^2 \alpha_b^{i_1 i_2} (0) + q_b^2 \alpha_a^{i_1 i_2} (0) \right] \bar{f}^2 (r_{ab}) \frac{r_{ab}^{i_1} r_{ab}^{i_2}}{r_{ab}^6}. 
\end{eqnarray}
Note that Eq. \eqref{eqn:12} includes the damping function (Eq. \eqref{eqn:11}) in order to take into account the unphysical divergences at small separations. To facilitate the calculation of $E_{disp}^{ind}$, the charges are obtained based on the L\"{o}wdin scheme \cite{Lowdin1950, Ertural2019}. This approach effectively subtracts the total charge population of a given atom from its number of valence-electrons using a symmetrically orthogonalized basis set for the assignment of electron densities \cite{Nelson2020}.

\subsection{Computational Details}

The WF-based scheme for computing dispersion energies is now applied to the molecular crystals from the X23 set and graphite, hBN, and \ch{MoS2} bulk materials taken as representative layered systems in which dispersive interactions are responsible for holding together their chemically inert monolayers. The layered materials have P63/mmc symmetry with P63mc hexagonal space group. Each constructed unit cell reflects the underlying symmetry and it is simulated utilizing the \textsf{Quantum ESPRESSO} ab initio package \cite{Giannozzi2009}. Graphite has AB stacking although AA stacking is also possible (Fig. \ref{fig:3}a) \cite{Bosak2007}. hBN is also composed of hexagonal atomic layers in which the B atoms are surrounded by N atoms. The preferred orientation for this material takes $\text{AA}^{\prime}$ stacking such that the B atoms in one layer are precisely on top of the N atoms in the adjacent layer (Fig. \ref{fig:4}a) \cite{Pease1952}. The transition metal dichalcogende monolayers in bulk \ch{MoS2} are also arranged in the most preferred $\text{AA}^{\prime}$ stacking patterns (Fig. \ref{fig:5}a) \cite{Wilson1969}. 

The DFT calculations are done utilizing GGA with Perdew-Burke-Ernzerhof (PBE) exchange - correlation potentials \cite{Perdew1996}. The norm-conserving, scalar-relativistic pseudopotentials used for the core-valence interaction of graphite and hBN. The  Projector Augmented Wave (PAW), scalar-relativistic pseudopotentials used for the core-valence interactions of the X23 set and \ch{MoS2}. The kinetic energy cutoffs for wave function expansions have been set to $70$ Ry. The kinetic energy cutoffs for charge density expansions have been set to $280$ and $420$ Ry for the norm-conserving and PAW pseudopotentials. All structures are optimized using the Broyden-Fletcher-Goldfarb-Shanno (BFGS) scheme until the total energy change is less than $10^{-8}$ Ry and residual forces are less than $10^{-4}$ Ry/Bohr. Gaussian smearing with broadening of $0.02$ Ry is used. For comparative purposes, dispersion energies are also calculated using all available schemes in \textsf{Quantum ESPRESSO}, including semi-empirical pairwise methods D2, D3(0), D3(BJ), XDM, and nonlocal density functionals vdW-DF1, vdW-DF2, optB88-vdW, optB86b-vdW, rev-vdW-DF2, vdW-DF3-opt1, vdW-DF3-opt2.

For each material, a suitable WF function basis interpolating the PBE-GGA DFT results is generated using the \textsf{WANNIER90} code \cite{Pizzi2020}, which is then used to interpolate each electronic structure. The representation of graphite and hBN contains 16 WFs with $s$ and $p$ orbital projections with frozen energy windows of 4 and 8 eV relative to Fermi level, respectively. For \ch{MoS2}, there are 28 WFs based on Mo $s, d$ and S $s$, $p$ orbital projections. The energy band structure together with the WF basis and computed WF spreads and centers, are then utilized to calculate the lattice polarizability based on linear response as captured in the Kubo formula \cite{Kubo1957}. This is followed by the computation of the vdW dispersion energy as captured in Eq. \eqref{eqn:6}. As a quantitative measure for the dispersive interactions, we compute the  lattice energy for materials in the X23 set as $E_{lat} = \dfrac{E_{s}}{Z} - E_g$, where $E_{s}$ is the total energy of each molecular crystal, $E_{g}$ is the total energy of the individual molecule in the gas phase of the material, and $Z$ denotes the number of molecules in the unit cell \cite{Reilly2013}. For the layered materials, we compute the interlayer binding energy per atom $E_b = \dfrac{E_{bulk} -  2E_{mono}}{N}$, where $E_{bulk}$ is the total energy of the bulk, $E_{mono}$ is the total energy of the individual monolayer placed in the vacuum of 30 \AA, and $N$ denotes the number of atoms in the unit cell of each material.

The polarizability of the lattice in Eq. \eqref{eqn:9} is computed based on the modified \textsf{Berry.f90} file in \textsf{WANNIER90} code. For this purpose, an in-house \textsf{Python} code was created fully incorporating the atomic positions, WF spreads and centers, as well as overall electronic structure. The code further determines the dipolar tensor ($A_{ab}^{i_1 i_2} (i \omega) = \alpha_a^{i_1 i_2} (i \omega) \delta_{ab}$) and damping function (Eq. \eqref{eqn:10}) for the full computation of the vdW dispersion energy (Eq. \eqref{eqn:6}). Atomistic charges $q_i$  are also calculated using the L\"{o}wdin scheme from the \textsf{LOBSTER} code by subtracting the gross population $GP_i$ of a given sort of atom from its number of valence-electrons $N_i$ \cite{Lowdin1950, Ertural2019}. \textsf{LOBSTER} is based on an analytic projection from DFT computations.  

\section{\label{sec:3}Results and Discussions}

\subsection{Molecular Crystals from the X23 set}
The computational scheme described above is now applied to calculate the energies and cell volumes of the molecular crystals as part of the X23 dataset \cite{Otero2012, Reilly2013, Mortazavi2018}. The dataset contains experimentally measured lattice energies and volumes, which are often used as reliable benchmarks in the development of new computational tools especially for dispersive interactions.

\begin{figure}[H]
    \begin{center}
    \includegraphics[width = 1 \columnwidth]{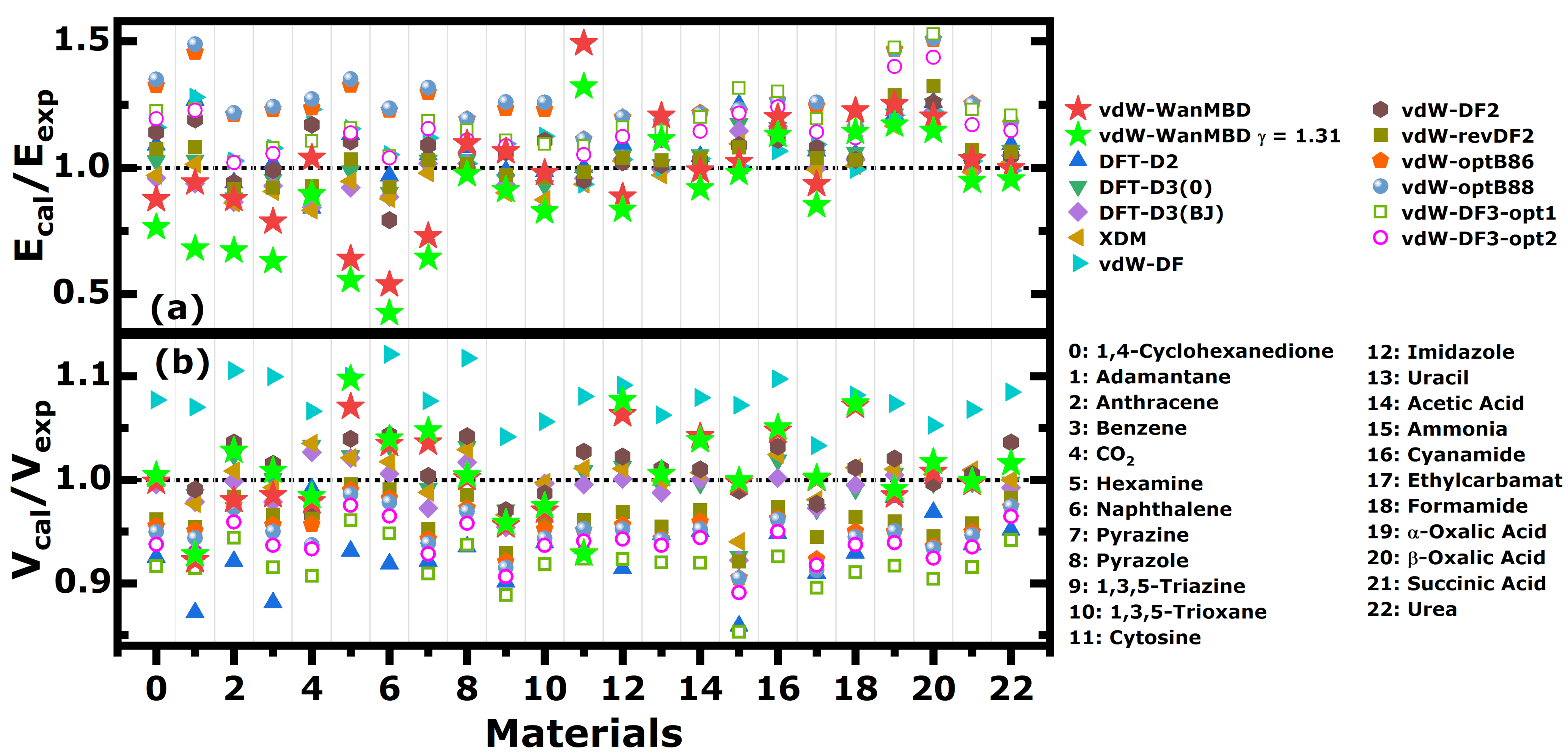}
    \caption{\label{fig:2} (a) Calculated lattice energy and (b) calculated cell volume normalized to the experimental values for each material in the X23 dataset as reported in Ref. \cite{Mortazavi2018} . The computational methods are denoted  next to panel (a) and the materials nomenclature is given next to panel (b). }
    \end{center}
\end{figure}

In Fig. \ref{fig:2}, we give the computed lattice energies and volumes using vdW-WanMBD with $\gamma={1; 1.31}$ coefficients in the damping function (see Eq. \ref{eqn:10}) and other vdW methods as implemented in Quantum ESPRESSO. Both properties are normalized to the experimental values for the different X23 materials. Fig. \ref{fig:2} shows that all methods are imperfect when compared with the experimental results. XDM, vdW-DF, and vdW-revDF2 show good agreement with the experimental energies for many materials, however there are substantial discrepancies with the experimental volumes. vdW-DF consistently overestimates the cell volume, vdW-revDF2 underestimates it, while XDM yields bigger and smaller than $V_{exp}$ volumes for the different materials. On the other hand, \textcolor{blue}{vdW-opt86 and} vdW-opt88 almost always yield larger energies, but smaller cell volumes when compared to experiments. For specific molecular crystals, such as Adamantine (No.1) and Naphtaline (No. 6), there are large energy and volume spreads. For example, vdW-opt86 and vdW-opt88 yield nearly 50$\%$ bigger than the Adamantane $E_{exp}$ energy, while $E_{cal}$ from vdW-WanMBD found with $\gamma=1$ is close to $E_{exp}$. vdW-WanMBD significantly underestimates the Naphtaline lattice energy for both $\gamma$ parameters, while vdW-opt86 and  vdW-opt88 give larger values. Nevertheless, Fig. \ref{fig:2} shows that for the majority of molecular crystals from the X23 dataset vdW-WanMBD result in lattice energies and cell volumes comparing reasonable well with reported experimental data.

\subsection{Layered Materials}

\subsubsection{The Case of Graphite}

\begin{figure}[H]
    \begin{center}
    \includegraphics[width = 1 \columnwidth]{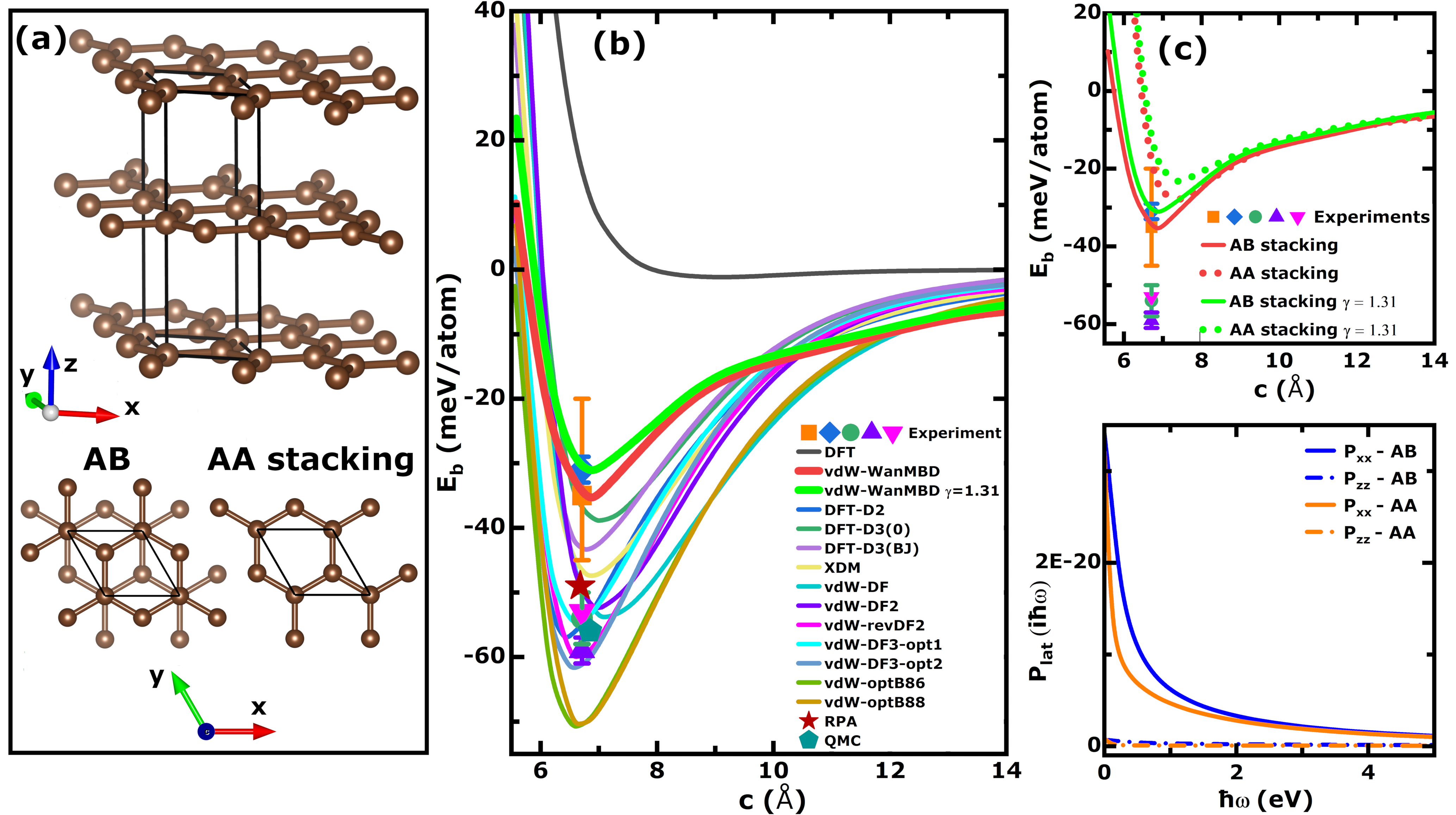}
    \caption{\label{fig:3}(a) Side view of graphite with the lattice constant $c$, which also serves as a measure of interlayer separation. Top views of AB- and AA- stacking are also shown. (b) Binding energy $(E_b)$ for the AB-stacking of graphite as a function of $c$. The PBE $+$ \texttt{vdW-WanMBD} results are compared with experimental data (orange from Ref. \citenum{Benedict1998}, blue from Ref. \citenum{Liu2012}, green and purple from Ref. \citenum{Rokni2020}, pink from Ref. \citenum{Tang2022}) and results from other computational methods, including from RPA \cite{Leconte2017} and QMC \cite{Spanu2009}. (c) Binding energy for the AB- and AA-stacking of Graphite as a function of $c$ with $E_{b0}^{AA} - E_{b0}^{AB} = 7$ meV/atom. (d) Lattice Polarizability $P_{lat}$ for the AB- and AA-stacking of graphite as a function of energy $\hbar \omega$.}
    \end{center}
\end{figure}

The computational scheme described above is now applied with the case of graphite. As C atoms within each layer experience strong covalent bonding, the interlayer interaction is much weaker and considered to be of vdW nature. The atomic structure of graphite is given in Fig. \ref{fig:3}a, which also shows the lattice parameter $c$ associated with the interlayer separation. Graphite layers organize with an AB stacking pattern, although an AA stacking configuration with a higher total DFT energy is also possible (Fig. \ref{fig:3}a). Since graphite is the non-polar material, only the vdW energy contributes into its dispersion energy  $E_b = E_b (DFT) + E_{disp}^{vdW}$. 

In Fig. \ref{fig:3}b, we show the computed interlayer binding energy for graphite in its most stable AB configuration. Results are given for the different methods as implemented in Quantum ESPRESSO together with our approach \texttt{vdW-WanMBD}. Standard DFT performs poorly in describing the structural stability of graphite; within the PBE level of approximation we find that the binding energy is $-0.82$ meV/atom with corresponding separation $c_0 = 8.10$ \AA. All other methods that take into account vdW interactions yield smaller equilibrium separation $c_0$ characterized by stronger interlayer binding energy $E_{b0}$ (the minimum point on each graph). Nevertheless, the collected results in Fig. \ref{fig:3}b fall into a wide range of values. Our \texttt{vdW-WanMBD} approach yields $E_{b0} = -36$ meV/atom with corresponding $c_0 = 6.83$ \AA \text{} for $\gamma = 1$ and $E_{b0} = -31$ meV/atom with corresponding $c_0 = 6.84$ \AA \text{} for $\gamma=1.31$. From all the pairwise methods, DFT-D3(0) has the highest energy $E_{b0} = -39$ meV/atom with corresponding $c_0 = 6.96$ \AA, while DFT-D2 has the lowest $E_{b0} = -57$ meV/atom with corresponding $c_0 = 6.39$ \AA. The results from the nonlocal density functionals tend to group towards stronger binding with a range outlined by the $E_{b0} = -53$ meV/atom, $c_0 = 6.97$ \AA \text{ } from vdW-DF2 and $E_{b0} = -71$ meV/atom, $c_0 = 6.58$ \AA \text{} from vdW-optB86. 

Experimental results measuring interlayer interaction energies in graphite are also available. One distinguishes between interlayer binding energy \cite{Benedict1998, Liu2012, Rokni2020, Tang2022} and exfoliation energy (required to remove one layer from the 3D graphite) \cite{Girifalco1956, Zacharia2004}. The reported measurements for the interlayer binding energy with their equilibrium separations also show significant discrepancies between each other, as given in Fig. \ref{fig:3}b. The different techniques used to find $E_{b0}$ and $c_0$ apparently show wide range in values, similar to the computations. Benedict et al. \cite{Benedict1998} extrapolated the binding energy of graphite to $-35 (+15, -10)$ meV/atom based on collapsed carbon nanotube measurements. Liu et al. measured binding energy for highly oriented pyrolytic graphite to be $31 \pm 2$ meV/atom \cite{Liu2012}. More recently, Rokni et. al reported $ -54 \pm 4$ and $-59 \pm 2$ meV/atom values based on an atomic force microscopy technique \cite{Rokni2020}. Tang et. al estimated the binding strength of graphite to be $-53$ meV/atom based on atomic force microscopy \cite{Tang2022}. Furthermore, the graphite binding energy computed via QMC simulations is $-56$ meV/atom \cite{Spanu2009}, while recent RPA calculations give $-49$ meV/atom \cite{Leconte2017}. Our \texttt{vdW-WanMBD} with $\gamma = 1$ and $\gamma = 1.31$ yield results that are in very close agreement with the experiments in Ref. \citenum{Benedict1998} and Ref. \citenum{Liu2012}, respectively. On the other hand, vdW-DF3-opt1 agrees very well with the data from Refs. \citenum{Rokni2020, Tang2022} (green and pink points), while vdW-rev-DF2 is in agreement with Ref. \citenum{Rokni2020} (purple point). Fig. \ref{fig:3}b further shows that vdW-opt86 and vdW-opt88 methods yield $E_{b0}$ that is $10$ meV/atom smaller than the largest experimental value $E_{b0} = -59$ meV/atom in Ref. \citenum{Rokni2020}. Also, DFT-D3(0), vdW-DF and vdW-DF2 tend to give larger equilibrium distances $c_0 = (6.96, 7.09)$ \AA \text{ }compared to the experimentally reported in Ref. \citenum{Bosak2007}.

Let us now compare the binding energy for the AB and AA stacking configurations as computed by the \texttt{vdW-WanMBD} method. Fig. \ref{fig:3}c shows that $E_{b0}^{AA}$ is $7$ meV/atom higher than $E_{b0}^{AB}$ for both $\gamma$. The equilibrium separations for AA graphite with $\gamma = 1$ and $\gamma = 1.31$ are found at $7.40$ \AA \text{} and $7.34$ \AA. It means that the lattice constants $c$ of AA graphite are longer than 0.52 \AA \text{} and 0.50 \AA \text{} compared to the ones of AB graphite for both $\gamma$. It appears that the stronger AB binding is accompanied by a shorter equilibrium separation as compared to the AA pattern. Within our method, the polarizability tensor is a key property in the vdW interaction, and we determine that the reason behind the AB preferred orientation is traced to the response properties of graphite.  In Fig. \ref{fig:3}d, the polarizability components $P_{ij}$ are shown for both configurations. Graphite is an anisotropic material as evident from the much larger response along the layers ($P_{xx}$ components) compared to the one perpendicular to the layers ($P_{zz}$ components). However, while $P_{zz}^{AB}$ and $P_{zz}^{AA}$ are very similar over the entire domain in imaginary frequency,  $P_{xx}^{AB}$ experiences a much steeper enhancement compared $P_{xx}^{AA}$ in the small frequency range $(< 2 \text{eV})$ with $P_{xx}^{AB} \sim 3 P_{xx}^{AA}$ at $\omega \rightarrow 0$. 

Our binding curves diverge slowly compared to other methods at the long distances which means that the atomic polarizabilities predicted from our method are larger than ones from previous approach. This is understandable since our method takes into account the effect of long-ranged neighbors on atomic polarizability while others only consider a limited range of neighboring atoms.

\subsubsection{The Case of hBN}

\begin{figure}[H]
    \begin{center}
    \includegraphics[width = 1 \columnwidth]{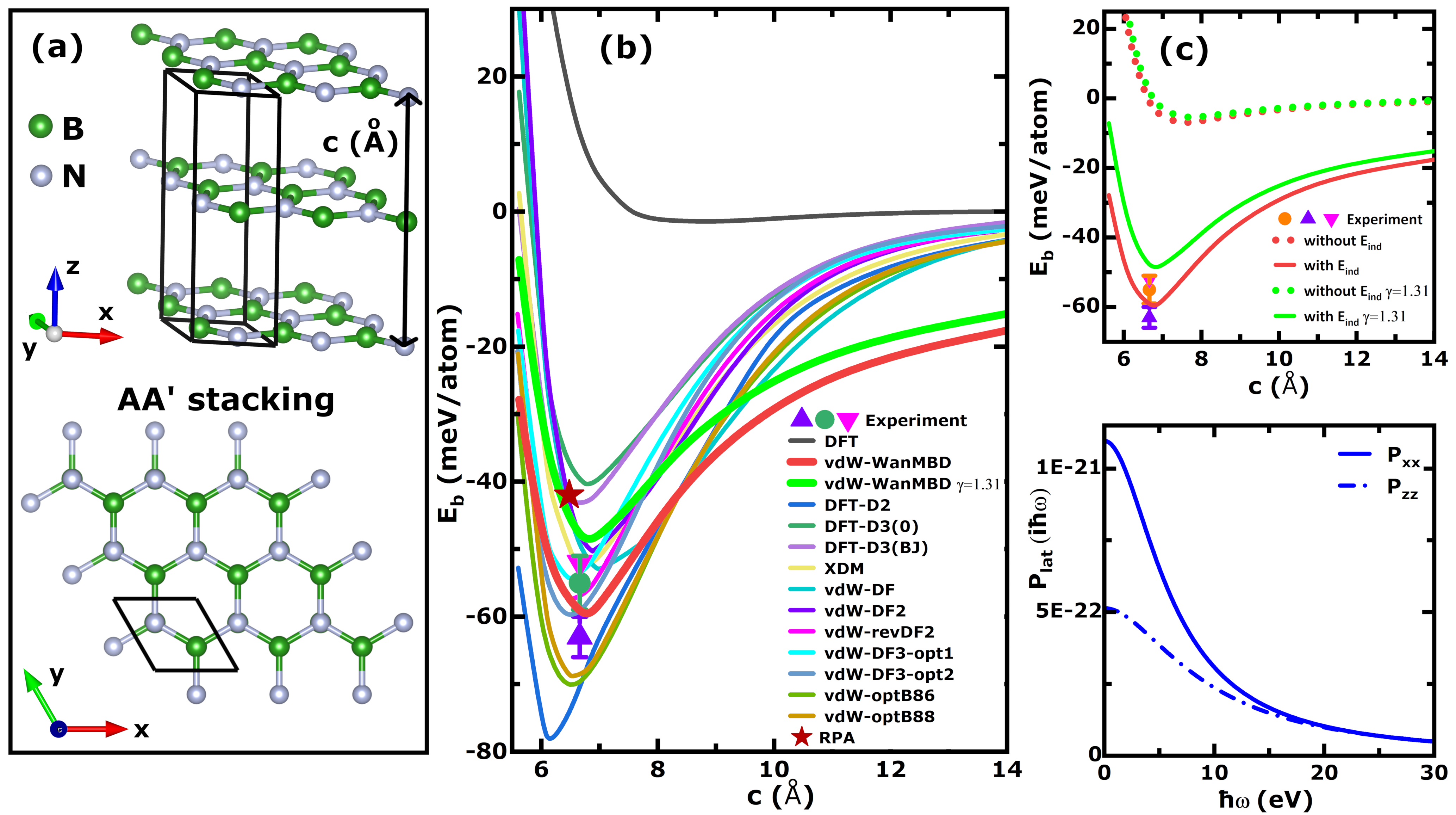}
    \caption{\label{fig:4} (a) Side view and top view of hBN bulk with $\text{AA}^{\prime}$- stacking with a lattice constant $c$ serving as an interlayer distance measure. (b) Binding energy $(E_b)$ for hBN bulk as a function of lattice constant $(c)$. The PBE $+$ vdW-Wan-MBD results are compared with the experimental data (green and purple from Ref. \citenum{Rokni2020}, pink from Ref. \citenum{Tang2022}) and resultrs from other computational methods, including from RPA \cite{Leconte2017}. (c) Binding energy for hBN bulk as a function of lattice constant $c$ with and without induction energies. (d) Lattice polarizability $P_{lat}$ for hBN bulk as a function of energy $\hbar \omega$. }
    \end{center}
\end{figure}

Let us now discuss the results for the interlayer binding energy in hBN in its most preferred $\text{AA}^{\prime}$ stacking configuration (Fig. \ref{fig:4}a) as a function of interlayer separation $c$. Fig. \ref{fig:4}b shows that standard DFT does not yield binding between the hBN layers. At the same time, vdW methods in Quantum ESPRESSO fall within a wide range of values between $E_{b0} = -40$ meV/atom with $c_0 = 6.74$ \AA \text{ }from DFT-D3(0) to $E_{b0} = -78$ meV/atom with $c_0 = 6.13$ \AA \text{ }from DFT-D2. Experimental data for hBN interlayer binding energy is also available. Rokni et al. show the cohesion energy of hBN crystals at room temperature with an average value of $-55 \pm 4$ meV/atom using normal force microscopy, while they also reported $-63 \pm 3$ meV/atom energies using shear force microscopy \cite{Rokni2020}. RPA calculations give binding energy of $-42$ meV/atom with a corresponding $c_0=7.71$ \AA \cite{Leconte2017}.

After applying our DFT-\texttt{vdW-WanMBD} approach as described in Eq. \eqref{eqn:6}, however, we obtain $E_{b0} = -7$ meV/atom at $c_0 = 7.40$ \AA \text{ }as given in Fig. \ref{fig:4}c. This so-obtained value is about an order smaller that the reported experimental results. To understand the reason behind this large discrepancy, we consider the polarization of the material. Similar to graphite, hBN is an anisotropic material with $P_{xx}$ being much larger than $P_{zz}$. However, hBN polarization is much smaller than the one for graphite, as can be inferred from Figs. \ref{fig:3}d and \ref{fig:4}d. Where is the “missing” contribution to the hBN binding energy? It is well known that hBN is a polar material whose bonds express themselves as uneven distribution of charge densities around the atoms. Specifically, the positively charged accumulation of electron density on B atoms and the negatively charged accumulation of electron density on N atoms create unequal sharing of electrons resulting in permanent dipoles in the material. This means that in addition to the vdW interaction originating from induced fluctuating dipoles, there is an induction interaction due to the permanent dipoles in the material. 

The interlayer binding energy is now computed by including the vdW and induction contributions, such that $E_b = E_b (DFT) + E_{disp}^{ind} + E_{disp}^{vdW}$. The results in Fig. \ref{fig:4}b show that $E_{disp}^{ind}$ has a significant effect bringing the equilibrium attraction to $E_{b0} = -61$ meV/atom with corresponding $c_0 = 6.62$ \AA  \text{ } for $\gamma = 1$ and $E_{b0} = -48$ meV/atom with corresponding $c_0 = 6.80$ \AA \text{ } for $\gamma = 1.31$. This energy found with $\gamma = 1$ is now in a very good agreement with available experiments reporting $E_{b0} = -55$ meV/atom in Ref. \citenum{Rokni2020} and $E_{b0} = -52$ meV/atom in Ref. \citenum{Tang2022}. The obtained interlayer energy with $\gamma = 1.31$ is underestimated compared to experimental values but comparable to vdW-DF2. The dramatic effect of the induction energy can be seen in Fig. \ref{fig:4}c, which shows that the vdW interaction actually plays a minor role and the interlayer interaction is essentially determined by the induction energy meaning that polarity of the B--N bonds in hBN cannot be neglected. 

\subsubsection{The Case of \ch{MoS2}}

\begin{figure}[H]
    \begin{center}
    \includegraphics[width = 1 \columnwidth]{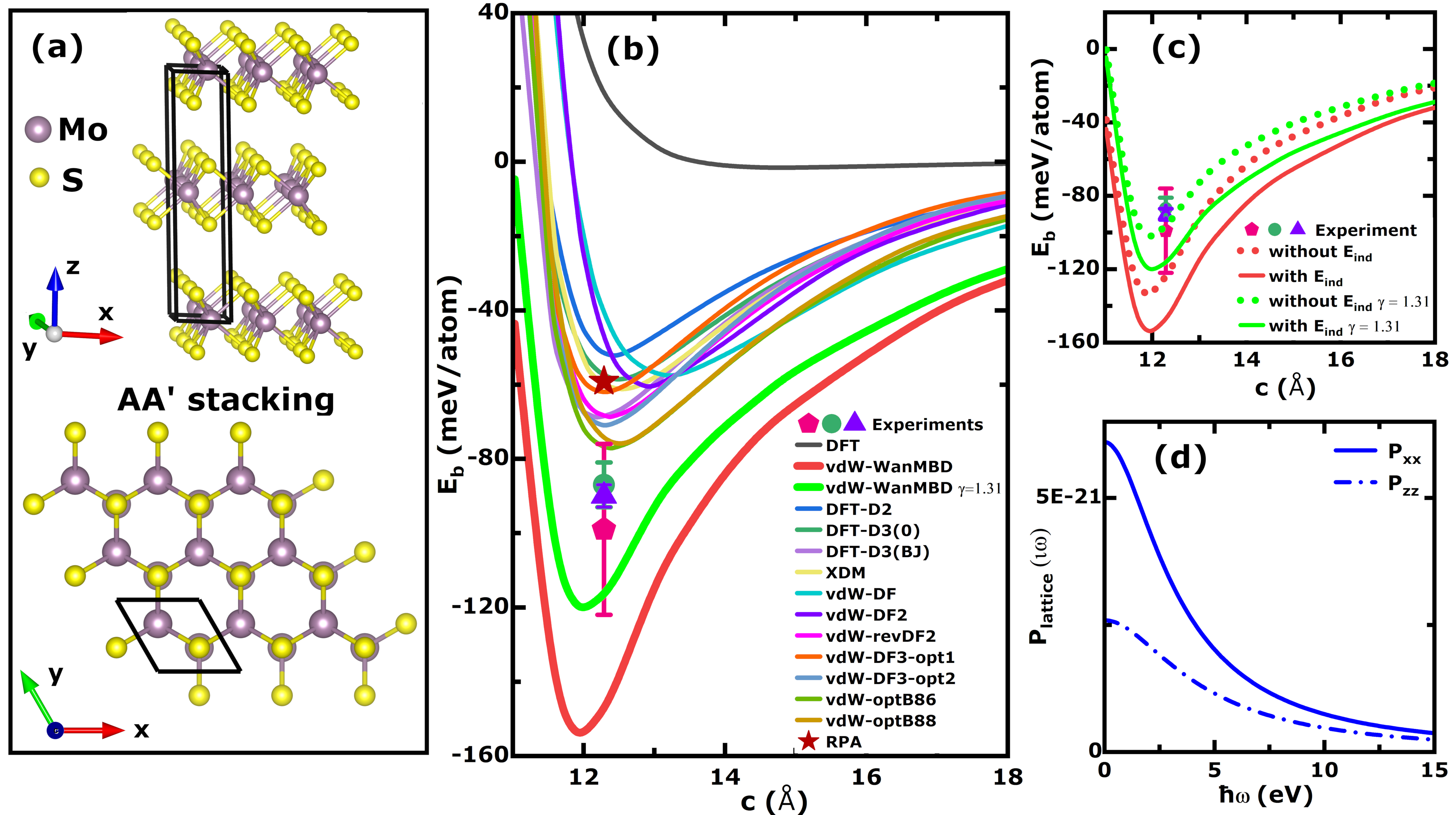}
    \caption{\label{fig:5}(a) Side and top view of \ch{MoS2} bulk with $\text{AA}^{\prime}$- stacking with a lattice constant $c$ taken as a measure of interlayer separation. (b) Binding energy $(E_b)$ for \ch{MoS2} bulk as a function of $c$. The PBE $+$ vdW-Wan-MBD results are compared with the experimental data (green and purple \cite{Rokni2020}, magenta \cite{Fang2020}) and results from other computational methods, including from RPA \cite{Gould2016}. (c) Binding energy of \ch{MoS2} as a function of lattice constant $c$ with and without induction energies. (d) Polarizability $(P_{lat})$ of \ch{MoS2} as a function of energy $(\hbar \omega)$.}
    \end{center}
\end{figure}

We also investigate the interlayer interaction in \ch{MoS2} in its most preferred $\text{AA}^{\prime}$ stacking configuration (Fig. \ref{fig:5}a). Similar to graphite and hBN, the stability of this material requires the inclusion of dispersive interactions since the DFT predicts very small equilibrium binding energy $E_{b0} = -1.6$ meV/atom at $c_0 = 14.5$ \AA \text{} (Fig. \ref{fig:5}b). Again, there is a wide range of energies as obtained from the different vdW schemes. While vdW-optB86 yields the lowest bindings $E_{b0} = -77$ meV/atom with $c_0 = 12.40$ \AA, DFT-D2 yields the highest value $E_{b0} = -52$ meV/atom with $c_0 = 12.42$ \AA, as shown in Fig. \ref{fig:5}b. For comparison, RPA gives the binding energy of $-59.09$ meV/atom with the corresponding $c_0= 12.29$ \AA as reported in Ref. \citenum{Gould2016}. The interlayer interaction is also computed within our \texttt{vdW-WanMBD} method, by evaluating the vdW energy as captured in Eq. \eqref{eqn:6}.  Since \ch{MoS2} is a polar material due to positive charges on Mo atoms and negative charges on S atoms \cite{Choudhuri2020}, the induction energy $E_{ind}$ must be included into the overall interlayer binding. 

To discern the role of induction, in Fig. \ref{fig:5}c we show results for $E_b = E_b (DFT) + E_{disp}^{vdW}$ and $E_b = E_b (DFT) + E_{disp}^{ind} + E_{disp}^{vdW}$ as a function of interlayer separation. It is found that for the first case and $\gamma=1$, $E_{b0} = -133$ meV/atom with $c_0 = 11.92$ \AA. For the inclusion of the induction energy, the computed interlayer binding energy is found as $-154$ meV/atom with corresponding $c_0 = 11.95$ \AA. For $\gamma = 1.31$, we find that $E_{b0}$ further decreases to $-122$ meV/atom with $c_0 = 11.98$ \AA. 

The available experimental results for interlayer binding energy in bulk \ch{MoS2} are also displayed in Fig. \ref{fig:5}b. Rokni et al. showed the binding energy of \ch{MoS2} crystals at room temperature with an average value of $-87 \pm 6$ meV/atom using normal force microscopy and $-90 \pm 3$ meV/atom using the shear force microscopy \cite{Rokni2020}. Fang et al. detetermined the interlayer binding energy of a mechanically exfoliated \ch{MoS2} to be $99 \pm 23$ meV/atom using in situ peeling-to-fracture method \cite{Fang2020}. One finds that the reported experimental values indicate a stronger interlayer binding energy when compared to the other computations while the opposite is true for our  \texttt{vdW-WanMBD} computation. The computted binding energy with $\gamma = 1$ shows stronger binding while, the one with $\gamma = 1.31$ is close the reported values from available experiments. 

\section{\label{sec:4}Conclusions and Outlook}

In summary, we have described a computational scheme for dispersive interactions that is based on a WF approach compatible with DFT simulations. This DFT-based \texttt{vdW-WanMBD} method captures naturally the electronic structure of the materials at the DFT level as well as its overall optical response. This method takes into account the many-body nature of vdW interactions and is a post-processing procedure with a low computational cost. Within the \texttt{vdW-WanMBD} method one can distinguish the different factors affecting dispersive interactions. 

As shown for the studied materials, there is a clear evidence that the full optical response plays a key role for the materials stability. By taking into account the layered materials anisotropy through explicit calculations of $P_{lat}^{ij}$, one can resolve the stability of different stacking patterns of a given material.  The approach we have presented further distinguishes contributions originating from fluctuating temporary and permanent dipoles. This is especially useful in the case of polar material, where the vdW energy associated with fluctuating dipoles and the induction energy due to permanent atomic dipoles are both responsible for the overall stability. 

The fundamental understanding of dispersive interactions is very complex and it is far from complete. As shown here and in other reports\cite{Tawfik2018,Bjorkman2012},  available computational methods give widespread equilibrium energies and distances for the X23 molecular crystals and the layered materials. The interplay of different assumptions, approximations, and input parameters can lead to sometimes vastly different results. Within the \texttt{vdW-WanMBD} method, we have shown this explicitly for the assumed Fermi damping function needed to avoid un-physical behavior at small separations. Using an input parameter $\gamma$ that scales the WF spreads can have strong effect in  the dispersive energy and/or lattice parameters, as is the case for several X23 materials and  MoS$_2$ (but it is less important for graphite and hBN). It is difficult to further quantitavley evaluate the scaling parameter for layered materials, since comparisons with experimental data is problematic. There is a scarcity of experimental data, and the reported data for binding energies in graphite, hBN, and MoS$_2$ seem to depend strongly on the measurement method. This is understandable since different defects, dopants, and impurities are measurement-dependent and they contribute to the relatively weak binding energy of layered materials. These external factors also affect the electronic structure and optical response thus a better coordination between measured and simulated materials is needed.

Our study demonstrates that the \texttt{vdW-WanMBD} method takes into account the overall electronic and optical properties of the materials and it can delineate different contributions for better understanding of factors important for dispersion interactions in materials. Similar calculations can be performed for different DFT functionals to examine how exchange-correlation effects influence the materials dispersive interactions. Future developments for taking into account beyond dipolar contributions may improve the performance of vdW-WanMBD when comparing with available data and help us better understand multipolar interactions.  This approach  can be viewed as a distinct and complementary method to other MBD and RPA frameworks. We hope this study can stimulate further experiments to measure interlayer interlayer interaction energies in layered systems for better validation of existing vdW computational schemes currently available in large scale DFT codes.

\begin{acknowledgement}
D.T-X.D. acknowledges support from Presidential Fellowship sponsored by University of South Florida. L.M.W. acknowledges financial support from the US Department of Energy under Grant No.DE-FG02-06ER46297. Computational resources are provided by USF Research Computing.
\end{acknowledgement}

\section*{\label{sec:6}Data availability}
The vdW-WanMBD code is available in figshare \cite{Woods2024} and Github \cite{Dang2024} platforms. The code contains a revised Fortran90 code and revised executable for the computation of various Berry phase properties from the Wannier90 package. It also contains a Python code for the computation of vdW, induction, and binding energies of layered materials using the WF basis representation as discussed in the paper. The data for the three materials studied here is also available. The Readme.txt file provides instructions of the computational procedure. 

\section*{\label{sec:7}Authorship contribution statement}
\begin{itemize}
\item[•] Diem Thi-Xuan Dang \orcidlink{0000-0001-7136-4125}: Methodology, Investigation, Data curation.

\item[•] Dai-Nam Le \orcidlink{0000-0003-0756-8742}: Validation, Methodology, Investigation, Formal analysis.

\item[•] Lilia M. Woods \orcidlink{0000-0002-9872-1847}: Writing – original draft, Supervision, Formal analysis, Conceptualization.

\end{itemize}

\section*{Declaration of Competing Interest}
The authors declare that they have no known competing financial interests or personal relationships that could have appeared to influence the work reported in this paper.

\bibliography{ref}

\end{document}